\def\bold#1{\setbox0=\hbox{$#1$}%
     \kern-.025em\copy0\kern-\wd0
     \kern.05em\copy0\kern-\wd0
     \kern-.025em\raise.0433em\box0 }
\def\slash#1{\setbox0=\hbox{$#1$}#1\hskip-\wd0\dimen0=5pt\advance
       \dimen0 by-\ht0\advance\dimen0 by\dp0\lower0.5\dimen0\hbox
         to\wd0{\hss\sl/\/\hss}}
\documentstyle[12pt]{article}

\newlength{\dinwidth}
\newlength{\dinmargin}
\newcommand{\resection}[1]{\setcounter{equation}{0}\section{#1}}

\begin{document}

\def\lq{\left [}
\def\rq{\right ]}
\def\LL{{\cal L}}
\def\VV{{\cal V}}
\def\AA{{\cal A}}
\def\MM{{\cal M}}
\def\eps{\epsilon}

\newcommand{\be}{\begin{equation}}
\newcommand{\ee}{\end{equation}}
\newcommand{\bea}{\begin{eqnarray}}
\newcommand{\eea}{\end{eqnarray}}
\newcommand{\nn}{\nonumber}
\newcommand{\dd}{\displaystyle}
\def	\frac		#1#2{{#1 \over #2}}
\def	\to		{\rightarrow }
\def    \pt	        {\mbox{$p_t$}}
\def\bc{\mbox{$B_c$}}
\def\mbc{\mbox{$m_{B_c}$}}
\def\fbc{\mbox{$f_{(b\to B_c)}$}}
\def\fbu{\mbox{$f_{(b\to B_u)}$}}
\def\fbi{\mbox{$f_{(b\to B)}$}}
\def\fps{\mbox{$f_{B_c}$}}
\def\ipb{pb$^-1$}
\def\eqnum#1{(\ref{#1})}       

\thispagestyle{empty}
\vspace*{2cm}
  \begin{Large}\begin{center}
{\bf $B_c$ Physics at Hadron Colliders}
\end{center}
\end{Large}
\normalsize
\vspace*{0.5cm}
\begin{center}
A.K.~Likhoded,~~S.R.~Slabospitsky\\
{\it Institute for High Energy Physics, Protvino, Moscow Region,
142284 Russia}
\end{center}
\vspace*{0.5cm}
\begin{center}
M. Mangano\\
{\it I.N.F.N., Sezione di Pisa,Italy\\
(CDF Collaboration)}
\end{center}
\vspace*{0.5cm}
\begin{center}
G. Nardulli\\
{\it Dipartimento di Fisica, Universita' di Bari\\
I.N.F.N., Sezione di Bari, Italy}\\
\end{center}
  \vspace{7mm}
\begin{center}
BARI-TH/93-137\\
Theory working group, UNK B factory workshop\\
Liblice Castle, Czechoslovakia, January 1993\\
\end{center}
\vspace*{15mm}
\begin{center}
  \begin{Large}
  \begin{bf}
  ABSTRACT
  \end{bf}
  \end{Large}
\thispagestyle{empty}
\begin{quotation}
  \vspace{5mm}
\noindent
In this paper we summarize the results of the theory working group
dedicated to the analysis of $B_c$ production at hadron colliders.
\end{quotation}
\end{center}
\newpage
\setcounter{page}{1}
\resection{Introduction}

Bound states of a $b$ and $\bar c$ quark pair (the $ B_c$ mesons)
 have never been
observed. They represent interesting objects for QCD
because their properties are expected to be calculable much in the
same way as properties of $c\bar c$ and $b \bar b$ states
have been studied theoretically in the last two decades by
making use, e.g., of potential models or QCD sum rules.

A first reason of interest in the study of $B_c$ physics arises from
the possibility to test quark potential models for systems made up by
quarks of different flavours as well as to compare these
predictions with other theoretical approaches. A second reason
of interest arises from $B_c$ decays. These decays are described at
the quark level by three classes of diagrams: annihilation
diagrams, c quark spectator  diagrams (e.g. $B_c \to \psi \mu \nu$
or $B_c \to \psi \pi$) and b quark spectator diagram (e.g. $B_c \to B_s X$).
The first class of diagrams would allow, in principle, to measure the leptonic
decay constant $f_{B_c}$. Note that this measurement would be enhanced at
least by the factor $|V_{cb}/V_{ub}|^2 \approx 10^2$ as compared to the
analogous measurement of $f_{B_d}$. Furthermore, the existence of two kinds of
spectator diagrams is a novel feature as compared to other
heavy meson decays.

For all these reasons, as well as to test QCD-like predictions
for $B_c$ production, among the future high energy high luminosity
hadron colliders a well defined place  should be reserved to the study of
$B_c$ properties, also because the production of such mesons at
these future machines is expected to be rather abundant.

\resection{The properties of $B_c$ mesons}

A possible approach
to the study of $b \bar c$ meson  properties: masses and  leptonic decay
constants, is provided by the the quark potential models
\cite{2,3,4,66,67}; another approach is represented by QCD sum rules
\cite {29}. In this Section  we review the predictions of both these
methods.

An extensive calculation of $b \bar c$ meson  properties has been performed
in \cite {4} by using the  Martin's potential \cite{12} :
\begin{equation}
V(r) = -8.064 + 6.869 r^{0.1}, \quad m_c = 1.8 \, GeV, \quad m_b
= 5.174 \, GeV,
\label{1}
\end{equation}
where the parameters of the potential are in GeV.

By solving the Schroedinger equation with the
potential (\ref{1}) one can obtain
the $B_c$ mass spectrum and the value of wave function of $B_c$  at
the origin. This method also predicts masses for radial and
orbital excitations; they are reported in Table 1 together with the
predictions of  \cite {2,3} that are based on flavour dependent
potentials. The values of Table 1 for the mass of the pseudoscalar
low lying state (the $0^-~ B_c$ meson) also agree with the result
of a relativistic potential model \cite {66} and with a QCD sum rules
calculation \cite {29}, within theoretical uncertainties. We take as an
indicative range of values for the $B_c (0^-)$ mass:
\begin{equation}
m_{B_c}~=~6250 \pm 100~ MeV.
\label{2}
\end{equation}

As for the mass difference between the $1^-$  and  $0^-$ states, potential
models give $m_{B^*_c}~-m_{B_c}\simeq 80 MeV$, which could be an
overestimate since $m_{B^*}~-m_{B}\simeq 50 MeV$. Indeed we should remember
that non relativistic potential models have been tested only for
the equal mass case and therefore, in the extrapolation to $\bar c b$
mesons, sizeable corrections could be introduced.

Note, that making use of $\Psi_{1s}(0)~=~0.369$~GeV$^{3/2}$ (see \cite{4}),
 one can calculate
the constant $f_{B_c}$ of weak decay of $B_c(0^-)$ meson :
$f_{B_c} = \sqrt{ \frac{12}{M} |\Psi (0)|^2} \simeq 570  MeV$
 where $f_{B_c}$ is defined by:
\begin{equation}
<0|{\bar c} \gamma_\mu \gamma_5 b|{B_c}(p)>~=~i f_{B_c} p_\mu~.
\end{equation}
On the other hand QCD sum rules \cite {29} give:
$f_{B_c} = 360 \pm 60 MeV$. We cannot solve this discrepancy at the moment
and we limit ourselves to quote a rather broad range of values for
$f_{B_c}$:
\begin{equation}
f_{B_c}~ =~ (300\div 600)~ MeV~.
\end{equation}

\resection{Hadronic production of $B_c$}

At the present there are some estimates of the cross
section for $B_c$ production
based on the parton picture and perturbative  QCD.
As a lower bound of the cross section one should consider the
$B_c$ meson pair production~\cite{21}.
\be
R_{B_c} = \frac{\sigma (B_c  \bar B_c)}{\sigma (b\bar b)}
\simeq (2\div 3)\cdot 10^{-4}\: ~~~~~(\sqrt{s}=100~ GeV).
\ee
The uncertainty in the prediction is mainly related  to the value of
$f_{B_c}$ $(=570~MeV$ in this calculation),
present in each vertex connecting $b\bar c$ with $B_c$.

For inclusive $B_c$ production (i.e. for $B_c \, \bar b \, c$ final state)
the above value of the cross section may be increased by about an order of
magnitude. Let us call \fbc the fraction of
$\bar b$ antiquarks that will evolve into a \bc\ meson.
No complete calculation of this parameter is available yet, but partial
estimates have been carried out \cite{21,22}. Reasonable estimates
of this fundamental parameter are in the range:
\be
\fbc=1\div 5 \times 10^{-3}~,
\ee
combining both perturbative and non-perturbative
contributions.
Other estimates based on the Monte Carlo codes (HERWIG, version 5.0)
are given in \cite{22}.

For the UNK energy range: $\sqrt{s} \sim 2$~TeV and with a luminosity:
${\cal L} \sim 10^{32}\:\mbox{cm}^{-2}\: \mbox{s}^{-1}$ the expected
annual yield of $B$-mesons is $\sim 10^{10}$, which,
according to our estimates, corresponds to $\sim  10^7 B_c$
produced per year.

\resection{$B_c$ decays}

{}From the study of inclusive decays of $B_c$ in the
framework of quark potential models \cite{24,25} one gets that the
contribution
to the total rate from different decay mechanisms  is  37\% from
$\bar c$-spectator decays, 45\% from $b$-spectator decays and 18\% from
$b\bar c$-annihilation. From QCD sum rules \cite {29} the corresponding
values are 48\%, 39\% and 13\% respectively.
The $B_c$ lifetime computed by
potential models is $\tau_{B_c}\simeq 5 \cdot 10^{-13}$~s,
i.e. $\Gamma_{tot}(B_c)\simeq 1.3\cdot 10^{-3}$~eV.
The corresponding values from QCD sum rules are
$\tau_{B_c}\simeq 9 \cdot 10^{-13}$~s,
i.e. $\Gamma_{tot}(B_c)\simeq 7.4 \cdot 10^{-4}$~eV.

Let us now consider specific decay channels.
A very clear signature for $B_c$ production  could be given by the decay
$B_c\to$ $\to J/\psi \mu^- {\bar \nu}_{\mu}$
(three leptons coming from the same
secondary vertex). The estimated branching ratio for this channel
is \cite{24,25,26,29}:
\be
BR(B_c\to J/\psi \mu^- {\bar \nu}_{\mu})\simeq (1\div 4)\cdot 10^{-2}~.
\ee
where the lower value corresponds to QCD sum rules prediction and the
upper one to potential models.

We can also consider hadronic decays, whose estimates follow
assuming the vacuum insertion approximation in effective nonleptonic
Hamiltonian. Predictions for these decays from different quark models
are summarized in Tables 2 and 3.

As a general characteristic, in $B_c$ decays the presence of $J/\psi$
in final states is rather frequent. The inclusive
$B_c\to J/\psi +X$ rate can be evaluated approximately in the
large $N_c$ limit, where the resulting branching ratio for
$B_c\to J/\psi + (light\: quarks\: or\: leptons)$
is 19\%$< Br(B_c\to J/\psi +X)<$24\%  \cite{24}.

In conclusion present theoretical investigations have not reached the
accuracy required to get very accurate predictions for the $B_c$ decays.
Table 4 summarizes predictions for the decay channels that are more
interesting from the experimental point of view.

\resection{Prospects for $B_c$ Discovery at CDF}

{}From the point of view of the detection, the two most important parameters
are the production rate and the branching ratios (BR) into accessible
decay modes. For the production rate we shall use the estimate for \fbc
reported in Section 3.
For the BR's we shall take the ranges of values reported in Table 4.
Notice that also the lifetime, as we have seen,
has a rather large range of values.
This is a critical parameter in view
of the possibility to reduce the background to the decay modes via the
presence of a secondary vertex.

In order to get a crude estimate of the possible signal at CDF, we will
normalize to the number of observed exclusive $B$ decays. The best
decay channel that allows full reconstruction is $\bc\to \psi\pi$. We will
compare this channel with the observed $B_u \to \psi K^+$, assuming equal
acceptance and reconstruction efficiency. Notice however that the efficiency
for the \bc\ decay is expected to be higher; in fact, the larger
mass of the \bc\ w.r.t the $B^+$ and the smaller mass of the pion w.r.t. the
kaon will give a larger impact parameter and a higher transverse momentum to
the decay pion.

Under the assumption of equal efficiency, we can write the following equation
for the number of expected reconstructed decays (the subsequent
$\psi\to\mu^+\mu^-$ decay is understood):
\be
\frac{(\bc \to \psi\pi^\pm)}{(B^+ \to \psi K^\pm)} =
\frac{\fbc}{\fbu} \times \frac{BR(\bc\to\psi \pi^\pm)}{BR(B^+\to\psi K^\pm)}
\approx 6 \; \fbc,
\ee
where we used the value of the BR given in Table~4 and \fbu=35\%.
Using the number of currently reconstructed $B\to \psi K^\pm$
(72 events in 9\ipb), we obtain $N(\bc\to \psi\pi^\pm) \approx 430 \fbc$. This
corresponds to 0.5--2 events, using the range of estimates for \fbc.
Considering the levels of the combinatorial background, this signal could be
detected with a larger integrated luminosity only in the upper range of \fbc,
unless the detection efficiency is significantly higher for this mode than for
$B\to\psi K$.

One could also hope to establish the presence of a \bc\ signal by looking at
the inclusive $\bc\to\psi\ell\nu$ decays, observing the presence of the third
lepton coming from the same secondary vertex as the $\psi$. In this case we can
write:
\be  \label{bctopsil}
\frac{(\bc \to \psi e(\mu) X)}{(B \to \psi X)} =
\frac{\fbc}{\fbi} \times \frac{2BR(\bc\to\psi\ell X)}{BR(B\to\psi X)}
\approx 20 \; \fbc.
\ee
This indicates that of the order of 2--10 \% of the $\psi$'s from $B$ decays
could be accompanied by an additional lepton ($e$ or $\mu$).
The requirement that this lepton come from the same vertex as the $\psi$ and
with a large \pt\  relative to the direction of the $B$ should reduce
significantly the possible background, but accurate feasibility studies are
still lacking. In equation~\eqnum{bctopsil}\ we used the Potential
Model (PM) estimate of the
inclusive  $\bc\to\ell\nu$ decay. The QCD Sum Rules (SR)
estimate would give a smaller, and
perhaps unobservable, signal.

\newpage
\begin{center}
\begin{Large}
\begin{bf}
Table Captions
\end{bf}
\end{Large}
\end{center}
\vspace{5mm}
\begin{description}
\item [Table 1]
Masses of $B_c$ mesons (in GeV) calculated using various potential
($\ast$ are the level masses, calculated without relativistic corrections).
\item [Table 2]
Two-body nonleptonic $b$-spectator decay rates (in units of
$10^{-6}$~eV), calculated in potential models:
BSW \cite {39} and ISGW \cite {49}. Values
of the parameters are:  $m_{B_c}=6.27$, $m_{B_s}=5.39$,
$m_{B^*_s}=5.45$, ,$m_{B}=5.27$, $m_{B^*}=5.33$, $m_{\pi}=0.140$,
$m_{\rho}=0.77$, $m_{K}=0.495$, $m_{K^*}=0.86$, $f_{\pi}=0.133$,
$f_{\rho}=0.216$, $f_{\omega}=0.195$, $f_{K}=0.162$, $f_{K^*}=0.216$~GeV.

\item [Table 3]
Decay rates (in units of
$10^{-6}$~eV) for some two body nonleptonic
$c$-spectator decays, calculated with ISGW \cite {49}
 form factors.
\item [Table 4]
Values of some parameters of interest for the decay of the \bc. The
expectations of different models (QCD sum rules (SR) or potential models (PM):
BSW \cite {39}; ISGW  \cite {49}),
together with uncertainty estimates, are included whenever available.
\end{description}
\newpage

\begin{table}
\begin{center}
{\bf Table 1}
\vspace*{0.5cm}
\begin{tabular}{|c|c|c|c||c|c|c|c|}
\hline
 State  & \cite{2} & \cite{3} & \cite{4} &
 State  & \cite{2} & \cite{3} & \cite{4} \\ \hline

 $1S \, \ast $ & 6.315  & -  & 6.301  &
 $1^1S_0  $ & 6.243  & 6.27  & 6.246  \\ \hline
 $2S \, \ast $ & 7.009  & -  & 6.893  &
 $2^1S_0  $ & 6.969  & 6.85  & 6.863  \\ \hline
 $3S \, \ast $ & -      & -  & 7.237  &
 $1^3S_1  $ & 6.339  & 6.34  & 6.329  \\ \hline
 $1P \, \ast $ & 6.735  & -  & 6.728  &
 $2^3S_1  $ & 7.022  & 6.89  & 6.903  \\ \hline
 $2P \, \ast $ & -      & -  & 7.122  &
 $1^1P_0 $ & 6.697  & -     & 6.645  \\ \hline
 $3P \, \ast $ & -      & -  & 7.395  &
 $1^1P_1  $ & 6.719  & -     & 6.682  \\ \hline
 $1D \, \ast $ & 7.008  & -  & 7.145  &
 $1^3P_1  $ & 6.740  & -     & 6.741  \\ \hline
 $2D \, \ast $ & -      & -  & 7.308  &
 $1^3P_2  $ & 6.750  & 6.77  & 6.760  \\ \hline
\end{tabular}
\end{center}
\end{table}

\begin{table}
\begin{center}
{\bf Table 2}
\vspace*{0.5cm}
\begin{tabular}{|c|c|c||c|c|c|}
\hline
 Mode & BSW & ISGW & Mode & BSW & ISGW \\ \hline
$B_c^+\to B_s +\pi^+$  & 47.8     & 67.7  &
$B_c^+\to B_s +\rho^+ $ & 19.2     & 31.  \\ \hline
$B_c^+\to B^*_s +\pi^+$ & 39.4     & 53.4  &
$B_c^+\to B^*_s +\rho^+$  & 177.8  & 234.  \\ \hline
$B_c^+\to B^+ +\bar K^0$  & 3.1    & 6.7  &
$B_c^+\to B^+ +\bar K^{*0}$  & 1.1  & 2.6  \\ \hline
$B_c^+\to B^{*+} +\bar K^0$  & 3.4    & 3.1  &
$B_c^+\to B^{*+} +\bar K^{*0}$ & 16.  & 18.  \\ \hline
$B_c^+\to B^0 +\pi^+$  & 1.49  & 2.9  &
$B_c^+\to B^0 +\rho^+$  & 1.45 & 3.3  \\ \hline
$B_c^+\to B^{*0} +\pi^+$ & 2.42 & 2.  &
$B_c^+\to B^{*0} +\rho^+$  & 13.6 & 12.  \\ \hline
$B_c^+\to B^+ +\pi^0$  & 0.05 & 0.1  &
$B_c^+\to B^+ +\rho^0$  & 0.05 & 0.12  \\ \hline
$B_c^+\to B^+ +\omega$  & 0.04 & 0.09  &
$B_c^+\to B^{*+} +\pi^0$  & 0.09 & 0.07 \\ \hline
$B_c^+\to B^{*+} +\rho^0$ & 0.48 & 0.48  &
$B_c^+\to B^{*+} +\omega$ & 0.39 & 0.38  \\ \hline
$B_c^+\to B_s + K^+$ & 3.35  & 5.  &
$B_c^+\to B^*_s +K^+$  & 2.6 & 3.9  \\ \hline
\end{tabular}
\end{center}
\end{table}

\begin{table}
\begin{center}
{\bf Table 3}
\vspace*{0.5cm}
\begin{tabular}{|c|c||c|c|}
\hline
$B_c^+\to \eta_c +\pi^+$  & 2.68  &
$B_c^+\to \eta_c +\rho^+$ & 6.2   \\ \hline
$B_c^+\to J/\psi +\pi^+$  & 2.75  &
$B_c^+\to J/\psi +\rho^+$ & 7.8   \\ \hline
$B_c^+\to \eta_c + K^+$    & 0.195 &
$B_c^+\to \eta_c + K^{+*}$ & 0.31  \\ \hline
$B_c^+\to J/\psi + K^+$ & 0.2  &
$B_c^+\to J/\psi + K^{*+}$  & 0.4  \\ \hline
\end{tabular}
\end{center}
\end {table}

\begin{table}
\begin{center}
{\bf Table 4}
\vspace*{0.5cm}
\begin{tabular}{|c|c|}
\hline
Lifetime & $0.5\div 1.5 \times 10^{-12}$ sec (PM) , $0.9\times 10^{-12}$ sec
	(SR) \\ \hline
\fps & $500\pm50$ MeV (PM) , $360\pm60$ MeV (SR) \\ \hline
BR($\bc\to\psi+X$) & $24\pm10$\% (PM) \\ \hline
BR($\bc\to\psi\ell\nu$) & $3\pm1$\% (ISGW,BSW) , $0.8$\% (SR) \\ \hline
BR($\bc\to\psi\ell\nu + X$) & $4.7$\% (ISGW,BSW) \\ \hline
BR($\bc\to\psi \pi$) & $0.2$\% (ISGW,BSW) \\ \hline
\end{tabular}
\end{center}
\end{table}

\end{document}